\documentclass[preprint,superscriptaddress,floatfix]{revtex4-1}

\usepackage{graphicx}
\usepackage{dcolumn}
\usepackage{bm}
\usepackage{amssymb}
\usepackage[utf8]{inputenc}
\usepackage{color} 
\usepackage{soul}  

\begin{document}

\title{Intra-Globular Structures in Multiblock Copolymer Chains from a Monte 
Carlo Simulation}

\author{K. Lewandowski}
\author{M. Banaszak}
\email[]{mbanasz@amu.edu.pl}
\homepage[]{http://www.simgroup.amu.edu.pl}
\affiliation{ Faculty of Physics,
A. Mickiewicz University \\
ul. Umultowska 85,
61-614 Poznan,
Poland
}

\date{\today}

\begin{abstract}
Multiblock copolymer chains in implicit nonselective solvents are studied by 
Monte Carlo method which employs a parallel tempering algorithm. 
Chains consisting of 120 $A$ and 120 $B$ monomers, arranged in three distinct 
microarchitectures: $(10-10)_{12}$, $(6-6)_{20}$, and $(3-3)_{40}$, collapse to 
globular states upon cooling, as expected. By varying both the reduced 
temperature $T^*$ and compatibility between monomers $\omega$, numerous 
intra-globular structures are obtained: diclusters (handshake, spiral, torus 
with a core, etc.), triclusters, and $n$-clusters  with $n>3$ (lamellar and 
other), which are reminiscent of the block copolymer nanophases for spherically 
confined geometries. Phase diagrams for various chains in the 
$(T^*, \omega)$-space are mapped. The structure factor $S(k)$, for a selected 
microarchitecture and $\omega$, is calculated. Since $S(k)$ can be measured 
in scattering experiments, it can be used  to relate simulation results to an 
experiment. Self-assembly in those systems is interpreted in term of 
competition between minimization of the interfacial area separating different 
types of monomers and minimization of contacts  between chain and solvent. 
Finally, the relevance  of this model to the protein folding is addressed.

\end{abstract}


\maketitle

\section{Introduction} 

A single multiblock copolymer chain in a nonselective solvent is an interesting
system to study because of its potential to form various nanostructures in a
globular state, such as: double droplet, lamellar, hand shake, spiral, and
disordered globule\cite{parsons}. 
Those nanostructures are reminiscent of  phases observed in block copolymer
melts\cite{hamley2004, woloszczuk_melt_with_solvent}, in general, and in
confined geometries\cite{im1, im14, im15, cylindrical_pores,
spherical_nanopores, sphericaly_confined}, in particular. Confinement can be 
one dimensional (thin films)\cite{im1}, two dimensional (cylindrical
pores)\cite{im14, im15, cylindrical_pores}, and three dimensional (spherical
pores)\cite{spherical_nanopores, sphericaly_confined}. The latter is the most
relevant analogue for a single polymer chain in a poor solvent, since the chain
collapses and tends to form a spherical globule.  Copolymer chains in spherical
confinement, with the interaction parameters similar to those considered in
this work, were studied,  using a coarse-grained model, 
in reference \onlinecite{spherical_nanopores}.  The following structures,
corresponding to those identified in multiblock copolymer chains, were
identified\cite{spherical_nanopores}: spheres with layers, helical-like or
handshake-like structures, and tricluster structures.

Ordering of copolymer    
nanostructures is driven by both lowering the temperature, which results in
decreasing the solvent quality, and lowering the compatibility between $A$ and
$B$, leading to a coil-to-globule transition and a subsequent segregation of
$A$ and $B$ monomers within the globule. There are interesting analogies: 
\begin{itemize}
\item the formation of a globule is a condensation of monomers, which is
similar to a gas-liquid transition,
\item segregation of $A$ and $B$ monomers resembles disorder-order and
order-order
transitions, resulting in  structures similar to those of copolymers in
confined geometries.
\end{itemize} 
At considerably lower temperatures the polymer chains crystallize, undergoing
both liquid-solid and
solid-solid transitions, with interesting packing effects,  as shown recently
for homopolymers \cite{Schnabel2009,Schnabel2009a,Seaton2010}.

An additional motivation for this study is its possible relation to structural
transformations in biopolymers\cite{simple_protein_models_review}. 
For example, the simplest models of proteins also employ only two types of
building blocks that is   hydrophilic, also referred to
as polar (P), and hydrophobic (H) \cite{random_heteropolymer_1,
random_heteropolymer_2}. These models, despite their excessive simplicity,
provide basic insight into formation of secondary and tertiary structures, but
in comparison to models with more types of monomers, they result in  less
cooperative folding and also in less designing
sequences\cite{simple_protein_models_review, two_letter_polymer_problem,
mulitple_degeneration_in_hp_model, Schiemann2005}. The sequence is referred to
as  designing when it can fold to the unique native state, that is the ground
state with the lowest energy. 
 
Majority of sequences in the HP model lead to degenerate native states, unlike
most proteins which exhibit 
the native state with fluctuations which probe various conformational
sub-states. Those sub-states are often very close to the native state, and are
caused by thermal fluctuations of atoms and slight displacements of amino acids
\cite{protein_native_states}.  
Increasing the number of types of monomers, with various interaction energies
(for example, as calculated in Ref. \onlinecite{miyazawa}), can improve these
models in terms of reproducing the properties of the real protein
systems\cite{many_letter_polymer_model_is_good}.
A  modification of the HP model, such as a change of the interaction energy
between H and P monomers, can also  improve this model noticeably
\cite{Schiemann2005}.  
In this work, we also use only two types of monomers, $A$ and $B$, but in
nonselective solvent, and this can provide some insight into ordering of
proteins with hydrophobic amino acids within the hydrophobic core.
It is worthwhile to reiterate that both monomers mimick the hydrophobic
behavior, at low temperatures, since the solvent is nonselective. 
 
The specific goals of the study are as follows:
\begin{itemize}
 \item to identify intra-globular structures of long multiblock copolymer
 chains with different compatibilities,
 \item to construct phase diagrams for those structures,
 \item to calculate structure factors for selected chains,   
 \item to relate results to protein folding problem.
\end{itemize}

\section{Model}
\subsection{Simulation box and environment}
Simulation is performed in a cubic box and the usual periodic boundary
conditions are imposed. The simulation box size is sufficiently large  for a
chain to fit in, and not to interact with itself across boundary conditions. We
simulate a single polymer chain and polymer-solvent interactions are included
in an implicit manner in polymer-polymer interaction
potential\cite{implicite_solvent}. This can be considered as a dilute polymer
solution.  

\subsection{Polymer model}
We use a coarse-grained model for  the polymer chain with  monomers of diameter
$\sigma$, taken also as the length unit. In this work, by ``monomer'' we mean
the basic building unit of the coarse-grained chain. Monomers are of two types:
$A$ and $B$. Neighboring monomers along the chain are connected via the bond
potential:
\begin{equation}
U_B(r)=\left\{
\begin{array}{ll}
\infty & \mbox{ for  } r < \sigma \\
0 & \mbox{ for  } \sigma \leq r \leq \sigma + \eta \\
\infty & \mbox{ for  } r > \sigma + \eta
\end{array}
\right.
\end{equation}
where $\sigma + \eta$ is the maximum bond length, and $\sigma + \frac{1}{2}
\eta$  is considered to be the average bond length.

Monomers that are not adjacent along the chain (nonbonded monomers), interact
via the  following square well potential:
\begin{equation}
U_N(r)=\left\{
\begin{array}{ll}
\infty & \mbox{ for  } r < \sigma \\
\epsilon_{ij} & \mbox{ for  } \sigma \leq r \leq \sigma + \mu \\
0 & \mbox{ for  } r > \sigma + \mu
\end{array}
\right.
\end{equation}
where $\sigma + \mu$  is range of the interaction potential,
$\epsilon_{ij}$ is interaction energy between monomers of types $i$ and $j$. We
assume that $\mu=\frac{1}{4}\sigma$, and
$\eta=\frac{1}{4}\sigma$\cite{fcc_offlatice}.

Chain bonds are not allowed to be broken, however they are allowed to be
stretched. Interaction parameter $\epsilon_{ij}$ is defined as:
\begin{equation}
  \begin{array}{c}
    \epsilon_{AA} = \epsilon_{BB} = -\epsilon  \\
    \epsilon_{AB} \in [-\epsilon; -0.1\epsilon] \\
  \end{array}
\end{equation}
The $\epsilon$ parameter, which is positive,  serves as an energy unit to
define the reduced energy per monomer $E^*/N$, and the reduced temperature
$T^*$ as: \begin{equation} \begin{array}{c} E^*/N =
\left(\frac{E}{\epsilon}\right)/N, \\ T^* = k_BT/\epsilon \end{array}
\end{equation} where $N$ is the number of chain monomers, and $k_B$ is
Boltzmann constant. Negative $\epsilon_{ij}$'s  indicate that there is an
attraction between monomers, and the presence of the solvent is taken into
account in an implicit manner\cite{implicite_solvent}.  By controlling the
relative strength of this attraction, via $T^*$, we effectively vary solvent
quality,  from  good to bad, which causes a collapse of the polymer chain, from
a swollen state to a globular state\cite{fcc_offlatice, fcc_simulation_model,
many_core}. The swollen and collapsed states are separated by the $\Theta$
solvent state, where the chain is   Gaussian.  This state is characterized by a
temperature $T^*_\Theta$, which is of the order of unity $T^*_\Theta \sim 1$
for this model.  Since we are interested in intra-globular structures, we
    mostly concentrate on temperatures below the $\Theta$-temperature.

Because we  vary $\epsilon_{AB}$'s in this work, a dimensionless parameter,
$\omega = -\epsilon_{AB}/\epsilon $ is introduced, which is a measure of
compatibility between A and B monomers. Lower values of $\omega$ mean lower
compatibility. As we increase $\omega$ from 0.1 to 1, we make the monomers
increasingly compatible, and for $\omega =1$ they become identical, converting
the copolymer chain into the homopolymer chain. 

\subsection{Polymer architecture} While we use only one chain length $N=240$,
with 120 $A$ monomers and 120 $B$ monomers, different multiblock
microarchitectures are considered: $(10-10)_{12}$, $(6-6)_{20}$, and
$(3-3)_{40}$. For the largest block size, $(10-10)_{12}$, the $A$ and $B$
monomers are expected to separate more easily at low temperatures, but for
smaller block sizes it may be more difficult to separate them into two phases
because of geometric frustrations. 

\subsection{Cluster count distribution} We differentiate between intra-globular
structures by the number of clusters of $A$ and $B$ monomers, and also by the
shape of those clusters.  We define cluster as a group of monomers of the same
type that are connected with each other either, directly or via a connected
path of monomers, and two monomers are connected if their distance is smaller
than the range of interaction potential.  We count the clusters in simulation,
and determine the equilibrium cluster count distribution (CCD) which is the
probability distribution for different counts of clusters. 

For each chain architecture and $\omega$,  CCD is calculated and plotted as a
function of $T^*$. These plots are used to determine  the phase diagrams in the
($T^*$, $\omega$)-space for each microarchitecture.  Since $n$-clusters, with
different $n$, can coexist for a given $(T^*, \omega)$, we show only the most
probable structures in phase diagrams.
 
\section{Method} We use the Metropolis\cite{metropolis} acceptance criteria for
the Monte Carlo (MC) moves. The MC moves are chain rotations, translations,
crankshaft rotations, and slithering snake moves. A Monte Carlo step (MCS) is
defined as an attempt to move once each monomer of the chain. 

Moreover, we use parallel tempering (replica exchange) Monte Carlo\cite{earl}
(PT) with feedback-optimized parallel tempering method\cite{feedback,
parallel_tempering_with_feedback} (FOPT). In the PT method $M$ replicas of
system are simulated in parallel, each at a different temperature $T^*_i$, with
$i$ ranging from 1 to $M$. After a number of MCS (in this work it is 100 MCS)
we try to exchange replicas with neighboring $T^*_i$ in random order with
probability: \begin{equation} p(T_i^* \leftrightarrow T_{i+1}^*) = \min[1,
\exp(-(\beta_i - \beta_{i+1})(U_{i+1} - U_i))] \end{equation} where $\beta_i =
1 / k_BT_i^*$ and $U_i$ is potential energy of replica at $T_i^*$.

Correctly adjusted  PT method allows a better probing of the phase space of the
system and prevents trapping in energy minima at low temperatures. Thus it
allows us to obtain better statistics in simulation and after a single
simulation we obtain results for the selected range of temperatures. We use
$M=24, 32$, and $40$ replicas.

A considerable challenge for the PT method is the selection of temperatures. As
we are interested in globular states, thus we choose mostly low temperatures
$T^* < T^*_\Theta$.  For chains $(3-3)_{40}$ and $(6-6)_{20}$ we used $T^*_i
\in [0.3; 1]$. For the $(10-10)_{12}$ chain we used $T^*_i \in [0.5; 1]$
because below $0.5$ nothing seems to change in the polymer structure. However
in each case for $\epsilon_{AB} \approx -\epsilon$ the highest value of $T^*$
was $1.1$ because coil-to-globule transition occurred for $T^* > 1$. The FOPT
method is used to obtain an optimized temperature set. In this method we start
with some temperature set (for example, linear or geometric) and  run the
simulation.   In the next iteration we obtain a more optimal temperature set. A
few iterations are required to obtain the optimized temperature set. This
method is described in detail in Ref. \onlinecite{feedback} and it was applied
to a polymer system in Ref. \onlinecite{parallel_tempering_with_feedback}, and
also in Ref. \onlinecite{matsen_pt}.

For each chain architecture we run five iterations of FOPT and we use the
obtained temperature set in simulations. Each iteration of FOPT and each
simulation consists of $5\times10^5$ MCS in athermal conditions (mixing) and at
least $4\times10^7$ MCS in thermal conditions. First $2\times10^7$ MCS are used
to equilibrate system and the rest is used to collect data.

\section{Results and discussion} \subsection{Chain with $(10-10)_{12}$
microarchitecture} First, we present the results for the $(10-10)_{12}$
multiblock chain with a compatibility parameter  $\omega = 0.1$. At high
temperatures this chain is in a coiled state, and it is expected to undergo the
coil-to-globule transition upon cooling. Moreover, due to a low compatibility
between  $A$ and $B$ monomers, some ordered $A$- and $B$-rich nanostructures
are also expected at low temperatures. 

Indeed, the temperature dependencies for energy $E^*/N$, the heat capacity
$C_v$, and the radius of gyration $R_g^2$ (in $\sigma^2$ units), shown in Fig.
\ref{chain_10_10_e0.1_en}, indicate the above orderings as $T^*$ is decreased.
In particular,  while $E^*/N$ does not change much from  high $T^*$'s to about
$T^* \approx 1$, for smaller $T^*$'s it decreases rapidly, and this corresponds
to a maximum in $C_v$, observed in Fig. \ref{chain_10_10_e0.1_en}(b). Also
$R_g^2$  exhibits the most rapid change in this temperature region. Judging
from  the above and the  position of the higher peak in $C_v$, we estimate the
coil-to-globule transition temperature, as approximately $T_{CG}^* \approx
0.85$. We observe that below $T_{CG}^*$, $R_g^2$ is almost constant, decreasing
slightly upon cooling.  In Fig. \ref{chain_10_10_e0.1_en}(c)  we can also
observe a $C_v$ maximum at about $T^* \approx 0.56$  and a ``bump'' in $C_v$ at
about $T^* \approx 0.75$. This is elucidated below in terms of the cluster
formation.

Representative snapshots of the $(10-10)_{12}$ multiblock chain are shown in
Fig. \ref{coil_to_globule}.  A variety of globular structures can be seen, with
the number of clusters decreasing upon cooling. This effect can be quantified
by a cluster count distribution (CCD) diagram for different $T^*$, as shown in
Fig. \ref{chain_10_10_e0.1_ccd}.  As the system is cooled down to a globular
state, the nanostructures with less clusters are more likely to be formed. 

For  $T^* \approx T_{CG}^*$ about 7- to 10-clusters are the most probable. The
corresponding structures are loosely packed, disordered globules, as shown in
in Fig. \ref{coil_to_globule}(b).  At $T^* = 0.75$ the most probable structure
is 4-cluster [Fig. \ref{coil_to_globule}(c)] with probability about $0.4$,
however tricluster [Fig. \ref{coil_to_globule}(d)] and 5-cluster are also very
likely to be seen with probabilities of about $0.3$ and $0.25$, respectively.
The remaining probability, $0.05$, is distributed for the 6-cluster and
dicluster structures.  Then for $T^* = 0.62$ we observe a maximum in CCD for
tricluster structures with probability $0.9$. This seems to correspond to the
minimum in $C_v$ for this $T^*$.  Finally, at $T^* = 0.56$ we see that
tricluster and dicluster [Fig. \ref{coil_to_globule}(e)] structures are equally
probable with probability $0.5$ (it corresponds to the second peak in $C_v$).
At lower $T^*$ we observe dicluster structures with probability approaching
$1$.

As noted above, for $T^* = 0.56$ we observe both a maximum in $C_v$ and a
transition from tricluster to dicluster structures in the CCD diagram. From
this observation we suggest that the $C_v$ bump  for $T^* \approx 0.75$ is
caused by structural changes from 6-cluster, through 5- and 4-cluster, to
tricluster in this region.

Next, we simulate multiblocks with gradually higher compatibility parameters,
from  $\omega =0.2$ to $1.0$. We show only representative CCD diagrams (Fig.
\ref{chain_10_10_e0.6_and_other_ccd}).  As we increase $\omega$, the
coil-to-globule transition occurs at gradually higher $T^*$'s. In Fig.
\ref{diagram_10-10} we show this transition for different $\omega$'s, and as it
is increased, transformations between intra-globular structures with different
number of clusters  shift to  higher $T^*$'s, in accordance with the
$T^*_{CG}$'s behavior.

In Fig. \ref{chain_10_10_e0.6_and_other_ccd}(a) we show CCD for $\omega = 0.6$.
For $T^*$'s between $0.6$ and $0.7$ probability of finding dicluster structures
increase noticeably (compare with Fig. \ref{chain_10_10_e0.1_ccd}) and reaches
about 0.3. For $\omega = 0.7$ [Fig. \ref{chain_10_10_e0.6_and_other_ccd}(b)]
and the same $T^*$ range, probabilities of finding dicluster and tricluster are
almost equal. Further increasing of $\omega$ [Figs.
\ref{chain_10_10_e0.6_and_other_ccd}(c) and
\ref{chain_10_10_e0.6_and_other_ccd}(d)] causes higher probability of finding
diclusters in this region with probabilities about $0.6$ and $0.9$ for $\omega
= 0.8$ and $1.0$, respectively.

This effect can be interpreted in terms of the site percolation problem
\cite{percolation}. For the highest compatibility $\omega = 1$, $A$ and $B$
monomers can freely mix, since energetically they are indistinguishable. In
this case the copolymer chain becomes a homopolymer and the clusters have no
physical meaning, but we analyze them formally in order to be consistent with
the rest of this work. The effect described in this paragraph is also relevant
to other $\omega$'s, but lower $\omega$'s,  monomers are less miscible.  At low
$T^*$'s, the monomers are densely packed spheres in a globule and most of them
have $Z=12$ nearest neighbors. Since the type of two neighboring monomers is
fixed (these are monomers along the chain - with an exception of the terminal
monomers which have only one type of fixed neighbor), only $Z=10$ neighbors of
varied types. Critical percolation threshold for face centered cubic lattice
(which can be used here since we have densely packed spheres in a globule) is
$p_c \approx 0.119$ \cite{percolation}. In our case, since we have 120 $A$ and
120 $B$ monomers, $p \approx 0.5$. Moreover in case of the $(10-10)_{12}$ chain
there are always blocks of 10 monomers of the same type, therefore in this case
it is sufficient that only one of 10 monomers from the block has contact with
another monomer of the same type in order to observe the percolation effect.
Therefore the dicluster structures (in which each monomer type creates a
continuous phase within globule) are very common.

For a better insight into dicluster structures we show representative snapshots
of these structures at $T^*=0.5$ (Fig. \ref{t0.5_10-10_structures}). As we
increase $\omega$ from $0.1$ to $0.5$ the structural changes are small [compare
Fig. \ref{coil_to_globule}(e) with \ref{t0.5_10-10_structures}(a)], then, as we
increase it further, handshake structures prevail. In Fig.
\ref{t0.5_10-10_structures}(b) we show a handshake structure and in Fig.
\ref{t0.5_10-10_structures}(c) the same structure without $B$ monomers. For
$\omega = 0.9$ we find many handshake structures and tori with a core, where
one type of monomers forms a torus around the other type [Fig.
\ref{t0.5_10-10_structures}(d)]. Then for $\omega = 0.95$ we find more
disordered structures, but very often  monomers of the same type seem to
aggregate [Fig. \ref{t0.5_10-10_structures}(e)]. Finally for  $\omega= 1$ we
observe disordered dicluster globules.
 
In Fig. \ref{diagram_10-10} we show  a phase diagram which collects data from
all $\omega$'s for the  $(10-10)_{12}$ chain.  We want to stress that
intra-globular structure is not exclusively defined by the number of clusters
and that presented phase diagram shows regions with the most probable
structures, but others are also present with smaller probabilities. 

\subsection{Structure factor} Since the CCD for a globule is not easily
measurable experimentally, we calculate structure factor (measured in
scattering experiments), $S(k)$, for the smallest compatibility $\omega = 0.1$.
We select $A$-monomers as the scattering centers and   calculate the isotropic
$S(k)$ as follows:

\begin{equation} S(k) = \frac{1}{N} \sum_{n,m} \frac{\sin{kr_{nm}}}{kr_{nm}}
\end{equation} where $k$ is a scattering vector length and $r_{nm}$ is a
distance between monomer $n$ and $m$. Results are presented in Fig.
\ref{sf_10-10}. 

For all $T^*$'s we can observe a small maximum at $k \approx 0.94$, which
corresponds to length of about $6.7\sigma$. Since it is also visible for
$T^*$'s above $T_{CG}^*$ we can conclude that it corresponds to the extended
size of a single block which is $10 \sigma$.

For $T^* = 0.7$ and $0.6$ we can see another small maximum for $k \approx 1.7$
which corresponds to length $3.7\sigma$. As we observe in Fig.
\ref{diagram_10-10}, the most probable structure for those $T^*$'s is
tricluster for which the thickness of the middle cluster  of the globule is
about $4\sigma$. Therefore, we think that this maximum is related to layers in
the tricluster structures. This maximum is not visible for $T^* = 0.5$. In this
temperature the most probable structure is dicluster and, since we ``scatter''
on $A$-monomers, we should obtain the structure factor of a flattened globule.

The above results indicate that the tricluster and dicluster structures may be
distinguished experimentally by the $S(k)$ measurements.

\subsection{Cluster count diagrams for other chain microarchitectures} The same
procedure, as described above, is performed for $(6-6)_{20}$ and $(3-3)_{40}$
chains and results are presented in Fig. \ref{diagram_6-6} and
\ref{diagram_3-3}.  In case of these chain microarchitectures it is necessary
to simulate them at lower $T^*$'s than those for the $(10-10)_{12}$ chain,
because many structural changes occur at lower temperatures.

First, we  discuss results for the $(6-6)_{20}$ chain. The coil-to-globule
transition occurs at a lower temperature, because  the energy variations are
smaller than those for the  $(10-10)_{12}$ chain. This is related to a higher
number of $A$-$B$ contacts along the chain, and the fact that those contacts
do not contribute effectively to the interaction energy.

In general, the regions of dominance for the 4-clusters and $n$-clusters (with
$n>4$) is similar to that for the $(10-10)_{12}$ chain, but it is different for
di- and triclusters. We can see that tricluster structures for $\omega < 0.6$
are the most probable at low temperatures, whereas for previously discussed
chain the dicluster structures prevail. This effect is explained in next
section in terms of the interfacial surface minimization.

For $\omega > 0.6$ we observe a behavior  similar to that of the $(10-10)_{12}$
chain. Monomers of different types mix and finally form disordered dicluster
within globules, as explained earlier in terms of the site percolation problem. 

Fig. \ref{t0.3_6-6_structures} presents selected snapshots of structures
observed for this chain. In Fig. \ref{t0.3_6-6_structures}(a) we see a
tricluster for $\omega = 0.1$ with flat $A$-$B$ interfaces. This structure
dominates over wide range of $\omega$'s. However, as we increase $\omega$'s
those tricluster structures become more spherical, for example, see Fig.
\ref{t0.3_6-6_structures}(d) for $\omega=0.7$, where the interfaces between $A$
and $B$ phases are more curved (like in handshake structures).  For the same
$\omega$, but at higher $T^*$'s, dicluster structures dominate, exhibiting
handshake structures, spiral structures [Figs. \ref{t0.3_6-6_structures}(b) and
\ref{t0.3_6-6_structures}(c)] and others.  

For $\omega \ge 0.8$ dicluster structures prevail in a wider $T^*$-range. Those
dicluster structures are very similar to those mentioned in previous paragraph.
Also a torus with a core appears here [Figs. \ref{t0.3_6-6_structures}(f) and
\ref{t0.3_6-6_structures}(g)]. 

Finally the $(3-3)_{40}$ microarchitecture is considered. The coil-to-globule
transition occurs in lower $T^*$'s (Fig. \ref{diagram_3-3}), because there are
more contacts between $A$ and $B$ monomers along the chain and therefore  lower
total energy.

Structures with a specified number of clusters occur in the same order as for
the other chain architectures, but they appear at lower $T^*$'s. For $\omega
\le 0.3$ the most probable structures consists of $5$ clusters arranged in
layers [Figs. \ref{t0.3_3-3_structures}(a) and \ref{t0.3_3-3_structures}(b)].
For $\omega  > 0.3$ at $T^* \approx 0.5$ we find 4-cluster structures and in
lower $T^*$ there is a region where 5-, 4-, and tricluster structures were
almost equally probable and, in our simulations, we could not identify the
dominating one.  For $\omega > 0.4$ the tricluster structures dominate and for
$\omega > 0.52$ dicluster structures dominate at low $T^*$'s. Again those
dicluster structures are mixed $A$ and $B$ monomers and they appear for similar
parameters (high $\omega$ and low $T^*$) as for the previously discussed
chains.

Figure \ref{t0.3_3-3_structures} presents snapshots of selected structures for
the $(3-3)_{40}$ chain. We notice that increasing $\omega$ yields more curved
structures, for example, the globule in Fig. \ref{t0.3_3-3_structures}(a) has a
rather flat interfaces between $A$ and $B$ phases, but in Fig.
\ref{t0.3_3-3_structures}(c) those interfaces are more curved.  As they become
more curved, they tend to connect with other layers. [see Figs.
\ref{t0.3_3-3_structures}(d)-\ref{t0.3_3-3_structures}(f)].

\subsection{Types of observed structures}

Structural changes observed in this work can be considered a result of the
competition between two effects:  minimization of  the $A$-$B$ interfacial
area, that is the number of contacts between $A$ and $B$, and  minimization of
the interfacial area between the chain and the solvent, that is the number of
contacts between the chain and the solvent.  Those effects are additionally
influenced by the chain microarchitecture.

If we consider a mixture of spheres of two incompatible types, we will observe
formation of two separate $A$- and $B$-rich phases. Adding bonds between
spheres frustrates the phase separation and this system tries to find a
structure, satisfying the chain microarchitecture constrains, with the minimum
number of $A$-$B$ contacts. For example, in symmetric diblock copolymer melts
one can observe lamellar structures with flat $A$-$B$ interfaces
\cite{woloszczuk_melt_with_solvent}. Asymmetric diblock
melts\cite{diblock_phase_diagram} or symmetric diblock melts with a
solvent\cite{woloszczuk_melt_with_solvent} form other nonlamellar structures,
such as: gyroid, cylinders, and spheres. 

On the other hand, in dilute polymer solutions, another minimization effect is
also significant.  Chains in a bad solvent minimize their contacts with solvent
and  form spherical globules, because sphere has the minimum surface area for a
given volume. For block copolymer melts, a spherical  confinement can be
introduced artificially. As shown in Refs. \onlinecite{spherical_nanopores} and
\onlinecite{sphericaly_confined} this   confinement can yield structures
similar to those presented in this study.

We assume, for the sake of this discussion, that the entropic effects are less
significant, and are not addressed.  The observed structures are thought to be
a result of a delicate interplay between the two above enthalpic effects.
Varying $\omega$'s changes the relative contributions of those two effects.
Increasing $\omega$ makes the first effect less significant. In the limit,
$\omega = 1$,  the first effect disappears (since $A$ and $B$ monomers are
fully miscible). From snapshots presented in this work, it can be seen that for
chains with higher $\omega$'s  globules are more spherical than those with
lower $\omega$'s.

As an example, we consider formation of tricluster structures in $(6-6)_{20}$
chain for $\omega=0.1$ at $T^* < 0.58$. Figure \ref{t0.3_6-6_structures}(a)
presents snapshot of a globule for those parameters. It is a tricluster
structure with clusters arranged in layers with almost flat $A$-$B$ interfaces.
This globule has an elliptical shape. Dicluster with a flat $A$-$B$ interface
seem to provide the smallest possible interfacial area, but to keep it flat, a
globule as a whole must be considerably flattened, what increases the contacts
between chain and solvent. On the other hand, compacting it into more spherical
shape increases the $A$-$B$ interface curvature. As a result one type of
monomers forms two separate clusters which order lamellarly. Increasing
$\omega$ results in forming more spherical globules and therefore in increasing
interface area between different types of monomers (see Fig.
\ref{t0.3_6-6_structures}), since, as described earlier, the $A$-$B$ penalty
for contacts is smaller for higher $\omega$'s.

Within-globular orderings can be related to the nanophase separations in block
copolymers, because a globule is locally dense system.  Since only symmetric
block sizes were used in this work, structures should correspond to bulk
symmetric diblock copolymers. In such systems lamellar phases are expected
(however gyroid, perforated lamellae, cylinders, and other structures were
recently found in sulfonated diblock copolymers with symmetric block sizes both
in experiment \cite{balsara_sulfonated} and in simulation
\cite{knychala_sulfonated}). Changing the $A$-$B$ interaction potentials should
only shift the order-disorder transitions temperature and it should not easily
lead to formation of nonlamellar structures. Why, therefore, we observe
handshake, spiral, and other nonlamellar configurations? As indicated earlier,
we consider it a result of a delicate interplay of two effects. As we increase
compatibility, the minimization of contacts between chain and solvent  becomes
a stronger effect than the minimization of the $A$-$B$ interfacial area, and in
order to form more spherical shapes of a globule, the $A$-$B$ interfaces tend
to be more curved. Thus, as a result we obtain the handshake, spiral, and other
structures.

We conjecture that asymmetric block sizes are necessary to obtain  gyroid-like
structures within a single globule. However they  were not considered in this
work. Probing different microarchitectures may lead to many other structures,
for example sufficiently long chains should reproduce a richer phase diagram
    with a cylindrical, gyroidal, or spherical intra-globular structures.

\subsection{Intra-globular structures and native states of proteins} This
coarse-grained model is probably too simplified to capture the complexity of
protein behavior, but it still may shed some light on it. For example,  in
order to create helical structures, hydrogen bonds are necessary
\cite{simple_protein_models_review}. Therefore, the spiral structures observed
in this study are not a manifestation of the helical-like structures found in
polypeptide chains, because our spirals are relatively thick--of the order of
few monomers in diameter [see Fig. \ref{t0.3_6-6_structures}(c)].

Amino acids can be roughly divided into two groups: hydrophobic and polar. Most
of simplified coarse-grained protein models capture mainly this property.
However, in our model solvent is nonselective and it becomes poor when we
decrease $T^*$. Therefore we can consider those globules as a hydrophobic cores
consisting of two types of amino acids. 

In this work we do not observe a single native state.  As can be seen in CCD
diagrams (Figs. \ref{chain_10_10_e0.1_ccd} and
\ref{chain_10_10_e0.6_and_other_ccd}) for a given $T^*$, structures with
different number of clusters can coexist. Only for the lowest $T^*$'s
probability of finding structures with a given number of clusters approaches 1.
But even in such case, there is degeneration of ``native'' states, because in
order to obtain it, chain can be folded in a variety of ways. For example, in
order to form dicluster structure like in Fig. \ref{t0.5_10-10_structures}(a)
it does not matter where different $A$ blocks are, as long as they are inside
the $A$ cluster.

However we would like to emphasize that the low compatibility between monomers
reduces the degeneracy of the inter-globular structures. There are less
possible ways to fold a chain into lamellar-like structure than into a
disordered globule.

\section{Conclusion} We  present an extensive Monte Carlo study of
intra-globular structures of long multiblock copolymer chains with alternating
blocks, using a discontinuous interaction potential.

Due to the chain architecture constrains, the interplay of minimization of the
$A$-$B$ interface with the minimization of the polymer-solvent interface yields
a rich phase diagram with a variety of intra-globular structures, such as
handshake, spiral, tricluster, torus with a core, lamellar, and many other
mixed and disordered structures. We relate our results to those for block
copolymer nanophases in bulk,  and find many similarities between them,
especially in spherical confinement. We also expect that other intra-globular
structures may be present in multiblock copolymer chains with asymmetric
blocks, such as analogs of gyroidal, cylindrical and spherical nanophases.

We analyzed the  $(10-10)_{12}$ chain behavior for various $T^*$'s and
$\omega$'s. From this analysis, and similar analysis of other chains
[$(6-6)_{20}$ and $(10-10)_{12}$], we construct phase diagrams in
$(T^*,\omega)$-space of the most probable $n$-cluster structures. In each case
decreasing $T^*$ leads to coil-to-globule transition, followed by transitions
between structures with $n$-clusters ($n$ decreases with $T^*$). The smallest
$n$ for low $\omega$ and low $T^*$ was 2, 3 and 5 for $(10-10)_{12}$,
$(6-6)_{20}$, and $(3-3)_{40}$, respectively.

From the structure factor of $(10-10)_{12}$ chain we can distinguish tricluster
and dicluster structures, and therefore it is possible to relate the numerical
predictions to an experiment.
 
Finally, we show that despite of the simplicity of this model, it still may
shed some light on highly complex behavior of proteins, for example: varying
compatibility between monomers within hydrophobic core may     reduce of the
degeneracy of the ground states.

\begin{acknowledgments} We gratefully acknowledge the computational grant from
the Poznan Supercomputing and Networking Center (PCSS) and grant N202 287338
from Polish Ministry of Science and Higher Education.  \end{acknowledgments}

\newpage \bibliography{bibliografia}

\newpage {\bf FIGURE CAPTIONS} \linebreak[1]

\begin{figure} \includegraphics*[width=8.6cm]{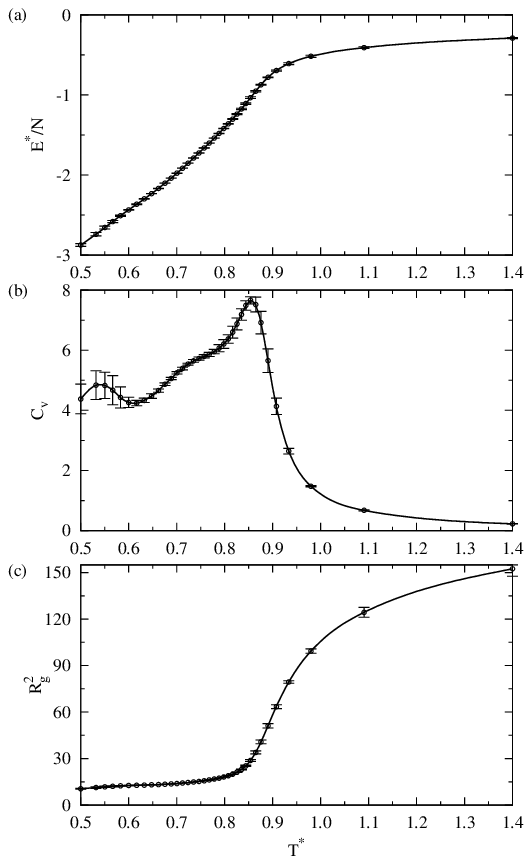}
\caption{\label{chain_10_10_e0.1_en} Lewandowski and Banaszak} \end{figure}

Fig. {\ref{chain_10_10_e0.1_en}: Results for the $(10-10)_{12}$ chain (error
bars are also shown): (a) reduced energy per monomer $E^*/N$, as a function of
reduced temperature $T^*$; (b) specific heat, $C_v$, as a function of $T^*$;
(c) squared radius of gyration, $R_g^2$, as a function of $T^*$.  }

\begin{figure} \includegraphics*[width=12.9cm]{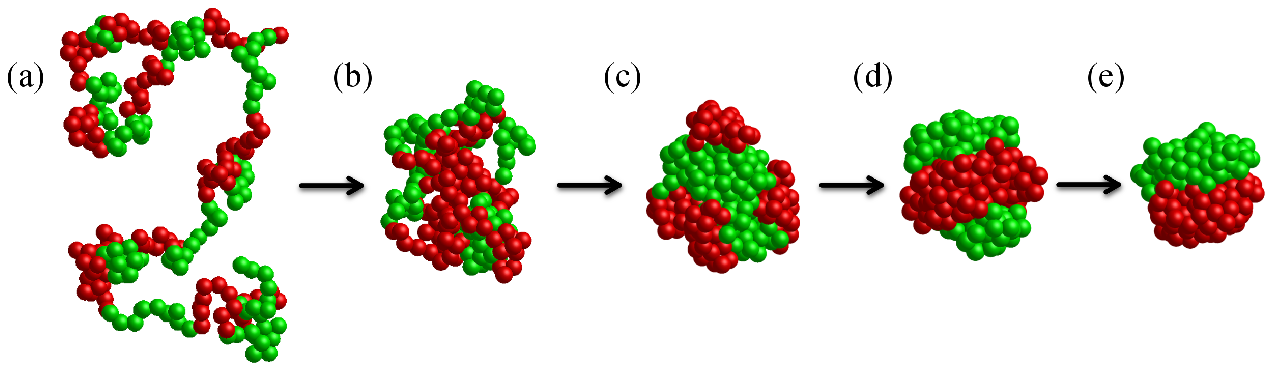}
\caption{\label{coil_to_globule} Lewandowski and Banaszak} \end{figure}

Fig. {\ref{coil_to_globule}: (Color online) Representative snapshots of the
$(10-10)_{12}$ chain with compatibility $\omega = 0.1$ in different
temperatures: (a) swollen state at $T^* = 1.2$, (b) coil-to-globule transition
at $T_{CG}^* = 0.85$, (c) 4-cluster structure at $T^* = 0.75$, (d) tricluster
structure at $T^* = 0.62$ and (e) dicluster structure at $T^* = 0.5$.  }

\begin{figure} \includegraphics*[width=8.6cm]{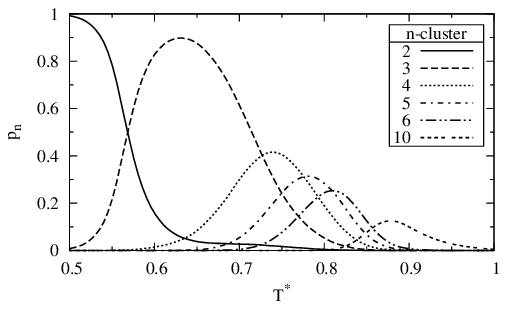}
\caption{\label{chain_10_10_e0.1_ccd} Lewandowski and Banaszak} \end{figure}

Fig. {\ref{chain_10_10_e0.1_ccd}: Probability of finding $n$-cluster structure
$p_n$, as a function of reduced temperature $T^*$, for the $(10-10)_{12}$ chain
with compatibility $\omega = 0.1$. For clarity, only selected $p_n$ lines are
shown.  }

\begin{figure} \includegraphics*[width=8.6cm]{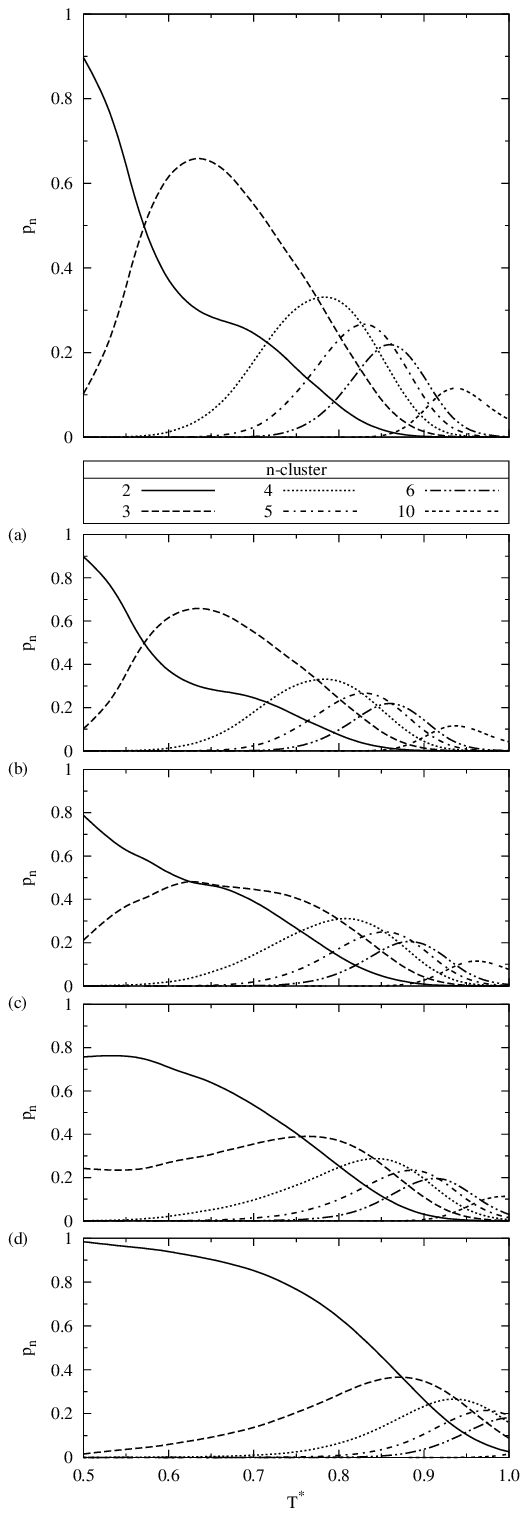}
\caption{\label{chain_10_10_e0.6_and_other_ccd} Lewandowski and Banaszak}
\end{figure}

Fig. {\ref{chain_10_10_e0.6_and_other_ccd}: Probability of finding $n$-cluster
structure $p_n$, as a function of reduced temperature $T^*$, for the
$(10-10)_{12}$ chain with compatibility: (a) $\omega = 0.6$, (b) $\omega =
0.7$, (c) $\omega = 0.8$, and (d) $\omega = 1.0$. For clarity, only selected
$p_n$ lines are shown.  }

\begin{figure} \includegraphics*[width=8.6cm]{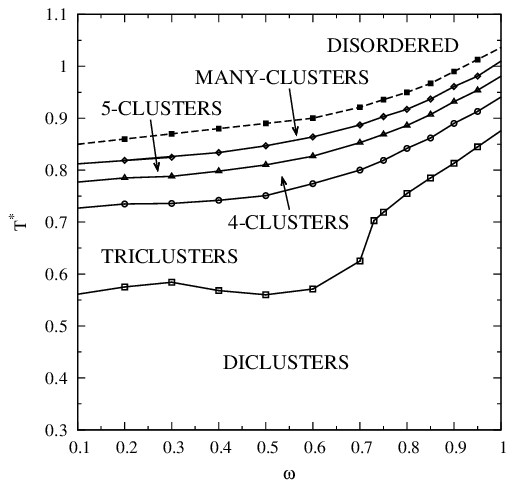}
\caption{\label{diagram_10-10} Lewandowski and Banaszak} \end{figure}

Fig. {\ref{diagram_10-10}: Phase diagram for the  $(10-10)_{12}$ chain in
$(T^*, \omega$)-space. Dashed line shows coil-globule transition. Solid lines
divides regions with the greatest probability of finding $n$-cluster
structures. Many-cluster region consists of structures with $n>5$ clusters.  }

\begin{figure} \includegraphics*[width=8.6cm]{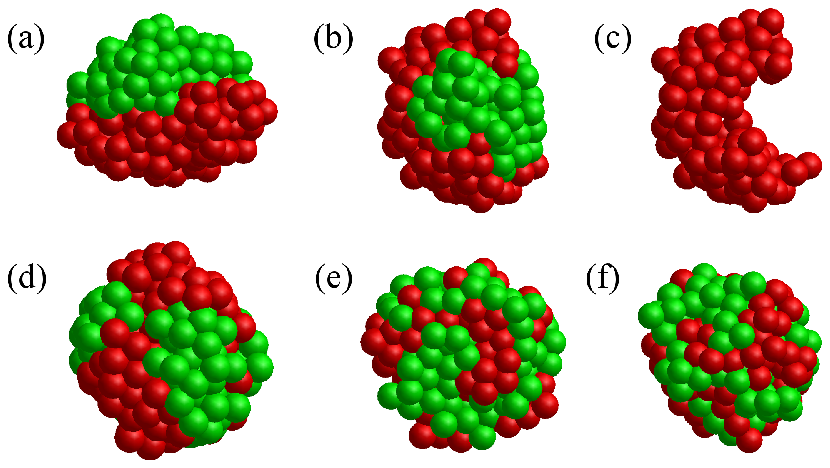}
\caption{\label{t0.5_10-10_structures} Lewandowski and Banaszak} \end{figure}

Fig. {\ref{t0.5_10-10_structures}: (Color online) Variety of dicluster
structures for the $(10-10)_{12}$ chain at $T^*=0.5$ for compatibility: (a)
$\omega = 0.5$, (b) $\omega =0.8$, (c) $\omega = 0.8$ (without $B$ monomers),
(d) $\omega = 0.9$, (e) $\omega = 0.95$, and (f) $\omega = 1$.  } 

\begin{figure} \includegraphics*[width=8.6cm]{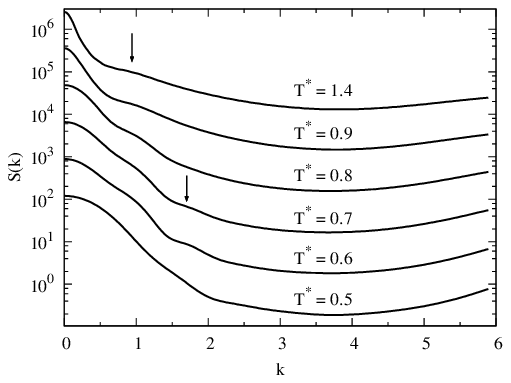}
\caption{\label{sf_10-10} Lewandowski and Banaszak} \end{figure}

Fig. {\ref{sf_10-10}: Structure factor $S(k)$ for the $(10-10)_{12}$ chain with
$\omega = 0.1$ at $T^* = 1.4$, $0.9$, $0.8$, $0.7$, $0.6$, and $0.5$. The
scattering profiles are offset vertically by factors of $10$, $10^2$, $10^3$,
$10^4$, and $10^5$, for clarity.  }

\begin{figure} \includegraphics*[width=8.6cm]{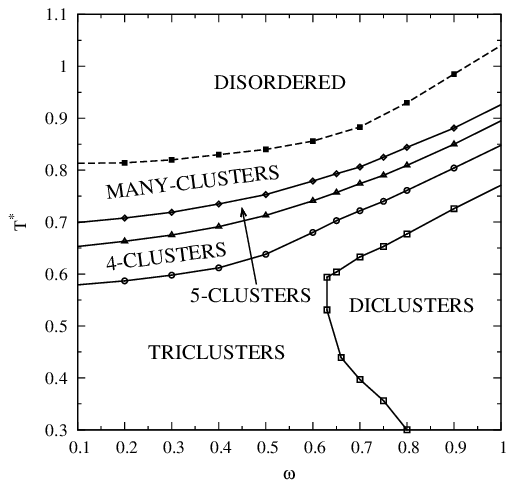}
\caption{\label{diagram_6-6} Lewandowski and Banaszak} \end{figure}

Fig. {\ref{diagram_6-6}: Phase diagram for  the $(6-6)_{20}$ chain in $(T^*,
\omega$)-space. Dashed line shows coil-globule transition. Solid lines divides
regions with the greatest probability of finding $n$-cluster structures.
Many-cluster region consists of structures with $n>5$ clusters.  }

\begin{figure} \includegraphics*[width=8.6cm]{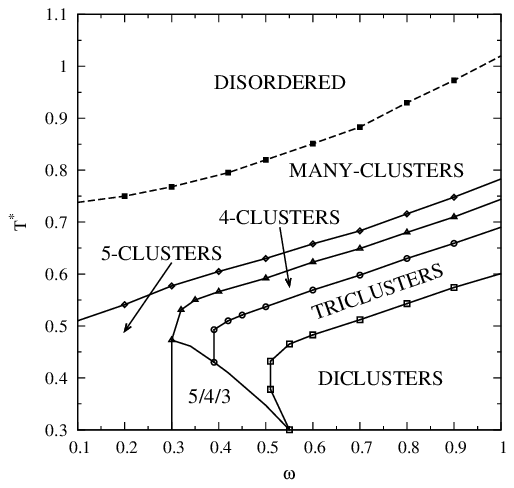}
\caption{\label{diagram_3-3} Lewandowski and Banaszak} \end{figure}

Fig. {\ref{diagram_3-3}: Phase diagram for the $(3-3)_{40}$ chain in $(T^*,
\omega$)-space. Dashed line shows coil-globule transition. Solid lines divides
regions with the greatest probability of finding $n$-cluster structures.
Many-cluster region consists of structures with $n>5$ clusters. In $5/4/3$
region probabilities of finding 5-, 4-, and triclusters is almost equal.  }

\begin{figure} \includegraphics*[width=8.6cm]{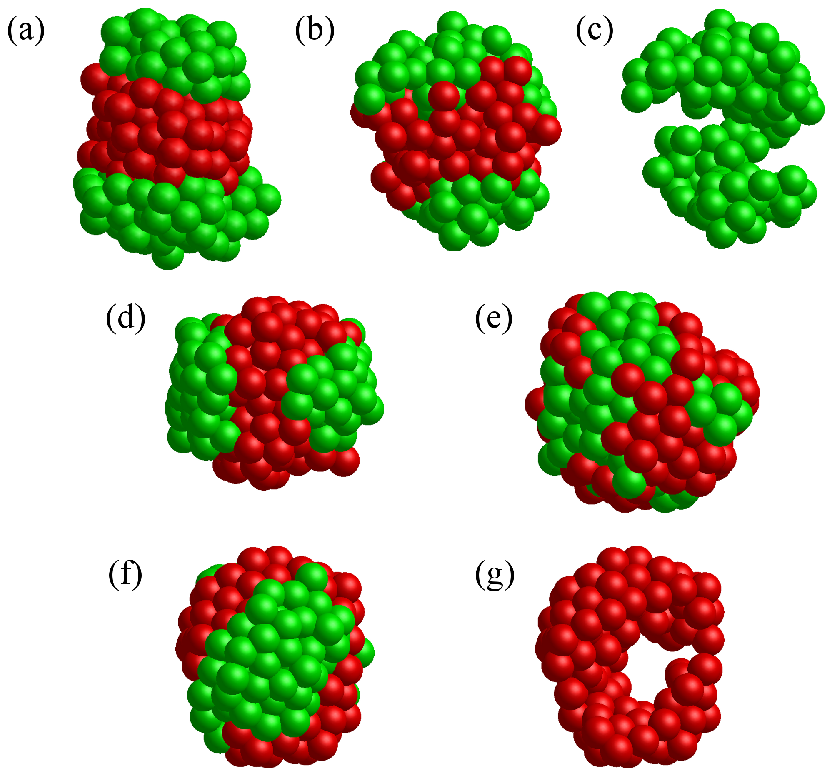}
\caption{\label{t0.3_6-6_structures} Lewandowski and Banaszak} \end{figure}

Fig {\ref{t0.3_6-6_structures}: (Color online) Variety of structures for the
$(6-6)_{20}$ chain at $T^*=0.3$ for compatibility: (a) $\omega = 0.1$
(tricluster), (b) $\omega = 0.7$ (at $T^*=0.5$, dicluster), (c) $\omega = 0.7$
(at $T^*=0.5$, without $B$ monomers, dicluster), (d) $\omega = 0.7$
(tricluster), (e) $\omega = 0.9$ (dicluster), (f) $\omega =0.8$ (dicluster),
(g) $\omega = 0.8$ (without $B$ monomers, dicluster).  } 

\begin{figure} \includegraphics*[width=8.6cm]{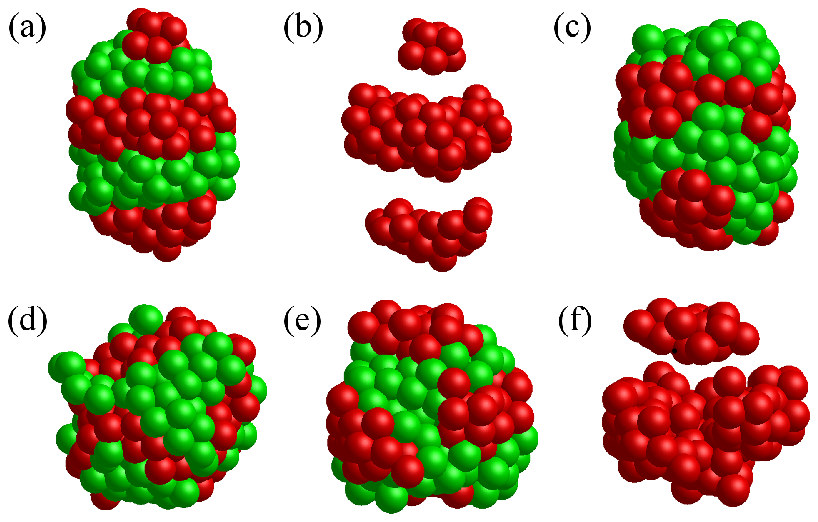}
\caption{\label{t0.3_3-3_structures} Lewandowski and Banaszak} \end{figure}

Fig. {\ref{t0.3_3-3_structures}: (Color online) Variety of structures for the
$(3-3)_{40}$ chain at $T^*=0.3$ for compatibility: (a) $\omega = 0.1$
(5-cluster, lamellar), (b) $\omega = 0.1$ (without $B$ monomers, 5-cluster,
lamellar), (c) $\omega = 0.31$ (4-cluster, lamellar), (d) $\omega = 0.5$ (at
$T^*=0.5$, tricluster), )e) $\omega = 0.51$ (tricluster), (f) $\omega = 0.51$
(without $B$ monomers, tricluster).  } 

\newpage

\end{document}